\documentclass[aps,prl,twocolumn,showpacs,amsmath,amssymb,floatfix]{revtex4}
\usepackage{graphicx}
\usepackage{dcolumn}
\usepackage{bm}
\usepackage{color}
\newcommand{\ba}{\begin{eqnarray}}
\newcommand{\ea}{\end{eqnarray}}

\begin{document}
\title{Manipulating Spins by Cantilever Synchronized Frequency Modulation:\\
A Variable Resolution Magnetic Resonance Force Microscope}
\author{K.C. Fong, P. Banerjee, Yu.\ Obukhov, D.V. Pelekhov and P.C. Hammel}\email{hammel@mps.ohio-state.edu}
\affiliation{Department of Physics,
             Ohio State University, 191 West Woodruff Ave., Columbus OH 43210}
\date{\today}
\begin{abstract}
We report a new spin manipulation protocol for periodically reversing the
sample magnetization for Magnetic Resonance Force Microscopy.  The protocol
modulates the microwave excitation frequency synchronously with the position of
the oscillating detection cantilever, thus allowing manipulation of the spin
magnetization independent of both magnetic field gradient strength and
cantilever response time. This allows continuous variation of the detected
sample volume and is effective regardless of spin relaxation rate. This
enhanced flexibility improves the utility of MRFM as a generally applicable
imaging and characterization tool.
\end{abstract}
\maketitle

Magnetic Resonance Force Microscopy (MRFM) is a ultra-sensitive technique for
studying and imaging small numbers of spins \cite{s:rmp, h:MagHandbook}. Since the original
proposal \cite{sidles91, s:prl}, MRFM sensitivity has steadily improved to the
single electron spin \cite{r:singlespin} and the thousand nuclear spin level
\cite{r:MaminStatisticalNMR, r:90-nmresolutionimaging}. Its superior spin sensitivity derives from large
magnetic field gradients that couple the magnetic resonance signal to low
noise, high quality factor $Q$ mechanical cantilevers \cite{sr}.  Because MRFM
has yet to be detected at the Larmor frequency, the force exerted by resonant
spins on the cantilever must be modulated at the cantilever frequency
(typically $f_c \sim 10$ kHz). It is advantageous to have the signal detection
bandwidth comparable to or greater than the inverse spin signal lifetime.

An approach to achieving the latter is to increase the cantilever response
bandwidth through active $Q$-damping \cite{Bruland:jap96, r:wago98,
r:microwireRFsource, Meier:LowGammaNuclei}. Although amplitude-based techniques
have advantages \cite{Budakian:FastCantileverPhaseReversals}, active
$Q$-damping requires excellent cantilever displacement detection sensitivity
\cite{Bruland:jap96, r:FeedbackCooling}. Alternatively, cantilever frequency
detection \cite{r:freqdetect:1991} such as implemented in the OSCAR (OScillating
Cantilever driven Adiabatic Reversals) \cite{r:hundredspin} protocol
circumvents this problem. OSCAR cleverly exploits the large magnetic field
gradient to manipulate the magnetization such that the resulting time-varying
spin force on the cantilever is manifested as shift in its frequency---a signal
detectable with a bandwidth limited only by noise considerations
\cite{r:freqdetect:1991}. However, to effectively invert the spin
magnetization, OSCAR requires field gradients of order $10^5$ T/m.

Here we introduce a new spin manipulation protocol based on frequency
detection. As in OSCAR, our protocol is effective regardless of the cantilever
response time. But, by frequency modulating the microwave frequency
synchronously with the cantilever oscillation, we efficiently detect the MRFM
signal regardless of the field gradient strength. This allows continuous
adjustment of the spatial resolution length scale, inversely proportional to
the field gradient, from coarse (requiring a low field gradient) to fine. This
capability is important for the development of MRFM as a flexible and broadly
applicable imaging and characterization tool. Frequency modulation excitation
\cite{r:nmr} has the additional advantage of minimizing spurious drive of the
cantilever. Here we apply it to spins with slow relaxation time $T_1\gg
1/f_c$), but it is also effective in the opposite limit \cite{h:oboukhov.dsp}.

\begin{figure}
\includegraphics[width=0.9\columnwidth]{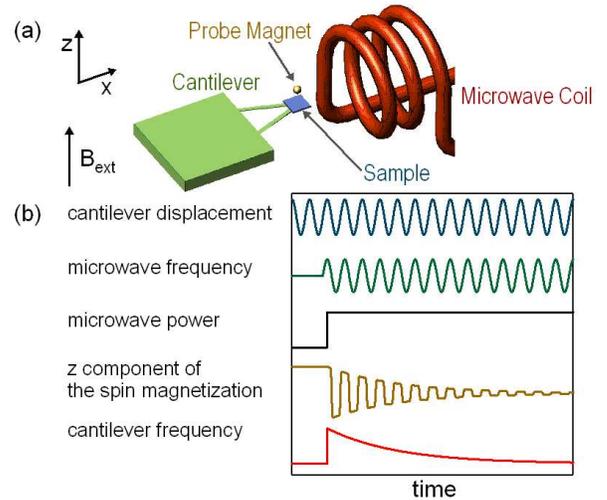}
\caption{(a) Schematic diagram of the MRFM setup. (b) Timing
diagram of the CASMO protocol. The microwave frequency is modulated in phase
with oscillation of cantilever. Following turn-on of the microwave power at an extremum of
the cantilever displacement, the spin magnetization is reversed periodically by
adiabatic fast passage. The resulting spin force on the cantilever modifies its
apparent compliance and hence its frequency; this constitutes the MRFM
signal.}
\label{fig:setup}
\end{figure}

The experiment (see Fig.~\ref{fig:setup}) was performed at $T = 4.2$ K on the
unpaired electron spins associated with E$^{\prime}$ centers, silicon dangling
bonds \cite{eaton:standardsample, castle:eprimecenters} (concentration $\sim
5\times 10^{17}$/cm$^3$) created in Clear Fused Quartz by 24.4 MRad $^{60}$Co
$\gamma$-irradiation \cite{WilmadSiliconDioxideSample}; at 4.2 K $T_1 = 2.3$ s.
The $100 \times 100 \times 5 \, \mu\rm m^3$ sample was glued on the end of a
silicon nitride cantilever (spring constant $k \sim 0.01$ N/m
\cite{Veeco:SiNsoft}); this lowers the cantilever frequency to 1.63 kHz. The
field gradient \(G_z = dB_z/dz\) is generated by a $\sim 40 \, \mu \rm m$
diameter spherical NdFeB micromagnetic probe, and the microwave frequency is
5.96 GHz.

Mechanical detection of magnetic resonance requires periodic modulation of the
force on the cantilever.  For slowly relaxing spins (relaxation time $T_1 \ll
f_c^{-1}$) the magnetic resonance technique known as adiabatic inversion is
used to periodically invert the spins at the cantilever frequency \cite{h:MagHandbook}.  The effective magnetic field $\vec B_{\rm{eff}}$ in the
rotating frame \cite{slichter} is given by
\begin{equation}
 \vec{B}_{\rm{eff}}(\vec{r}) =
 \vec{B}_{\rm{probe}} (\vec{r}) + \left(B_{\rm{ext}}
 - \frac{\omega_{\rm{rf}}}{\gamma}\right)\hat{z}+H_1\hat{x}
\label{eqn:effective_field}
\end{equation}
where $\gamma$ is the gyromagnetic ratio, $\omega_{\rm{rf}}$ and $H_1$ are the
frequency and magnitude of the transverse oscillating field, and $B_{\rm{ext}}$
and $\vec{B}_{\rm{probe}}$ are the applied and micromagnetic probe fields
respectively. If the effective field is rotated sufficiently slowly compared to
$\gamma B_{\rm eff}$, the spins follow the direction of the effective field by
adiabatic fast passage \cite{slichter}. MRFM experiments rely crucially on
rotating spins through manipulation of $\vec B_{\rm eff}$.

Our spin manipulation protocol inverts $\vec B_{\rm eff}$ through the second
term on the RHS of Eq.~(\ref{eqn:effective_field}), that is, by modulating the
microwave frequency $\omega_{\rm rf}$ synchronously with the cantilever
oscillation; we dub it CAntilever Synchronized frequency MOdulation (CASMO). As
shown in figure \ref{fig:setup}(b), the cantilever, driven by a positive feedback circuit,
oscillates continuously at its natural frequency, while the microwave frequency
is modulated such that the spin magnetization oscillates in synchrony with the
cantilever. The protocol is implemented through a digital signal processing program\cite{h:oboukhov.dsp}.

Microwave irradiation, modulated so as to periodically invert the spin
magnetization by means of cyclic adiabatic inversion, is initiated at an
extremum of cantilever oscillation.  This modulation is timed to ensure that
the spin magnetization, and hence the force exerted on the cantilever due to
the probe field gradient, is precisely in phase with the cantilever position
$z(t)$: $F_{\rm spin}(t) \propto z(t)$.
As a consequence, this force is evident as an addition to the apparent
cantilever compliance, \(\Delta k = F_{\rm spin}/z_{\rm pk} \) where $z_{\rm
pk}$ is the amplitude of the driven cantilever oscillation, and hence as a
change in the cantilever frequency \cite{Berman:OSCAR}. To thoroughly reverse
the effective field, we need a large FM deviation \(\Omega_{\rm{dev}} \gg
\gamma H_1/2\pi\). This is the limiting requirement that is traded for the
relaxed field gradient requirement encountered in OSCAR.

\begin{figure}
\includegraphics[width=0.8\columnwidth]{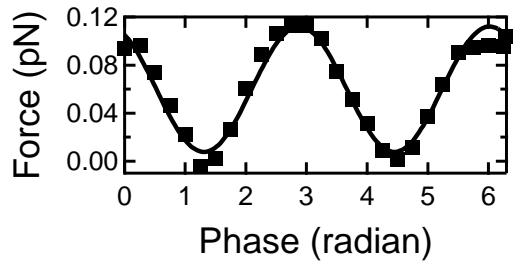}
\caption{Dependence of the MRFM signal on the phase of the microwave frequency
modulation relative to the cantilever position. The MRFM signal varies
periodically with phase because only the component of the force in phase with
the oscillating cantilever's position contributes to the signal.}
\label{fig:fm_phase}
\end{figure}

It is essential to control the phase of the FM with respect to the cantilever
oscillation correctly.
The oscillatory force is phase synchronous with the FM, and the cantilever
frequency only reflects the component of the force in phase with the cantilever
oscillation. Fig.~\ref{fig:fm_phase} shows the dependence of the CASMO
generated signal on the phase difference between the FM and the cantilever
oscillation. The MRFM signal is maximized when the phase difference is a
multiple of $\pi$, as expected. The slight offset results from a phase shift
added by the bandpass filter applied to the cantilever displacement signal.

\begin{figure}
\includegraphics[width=\columnwidth]{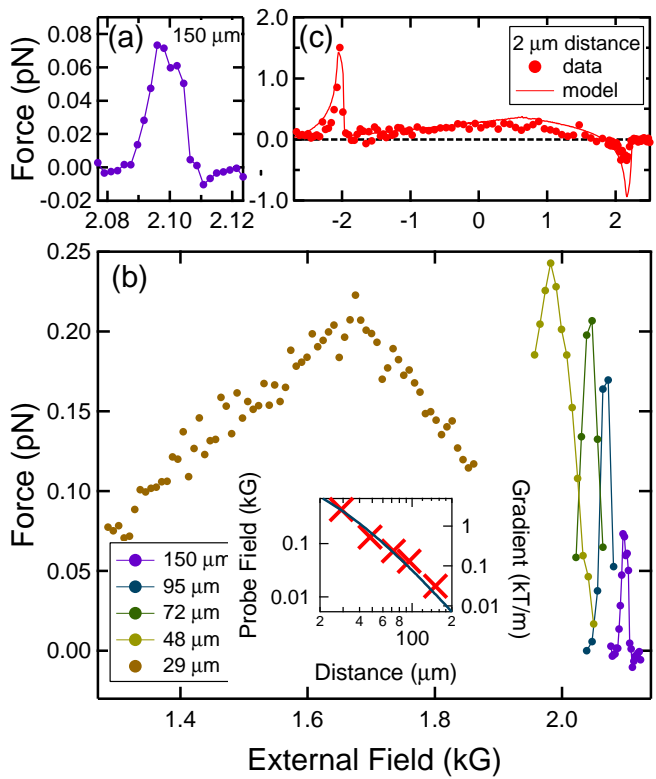}
\caption{Evolution of spectra with increasing field gradient: all panels show
the variation of the MRFM signal with external field. (a) At a probe-sample
separation $d = 150 \, \mu {\rm m},\, G_z \sim 30$ T/m ($G_z \Delta z \sim 1.5$
G), so the spectral width $\Delta H \sim 15$ G  is dominated by the frequency
modulation. (b) Decreasing $d$ increases $G_z$ and hence broadens the spectra $\Delta H \sim G_z \Delta z$ at
large $G_z$. Inset: Red crosses show the measured probe fields determined from the
MRFM signals, while the solid line is the calculated field and gradient from a
uniformly magnetized spherical magnet 44 $\mu$m in diameter. (c) For $d= 2 \, \mu$m,  $G_z \sim 10^5$ T/m dominates the width
$\Delta H > 4$ kG. The two peaks near  \(\omega_{\rm rf} /\gamma = \pm 2.1\)
kG, are the zero-probe-field resonance (ZPFR) signals arising from remote
regions of the sample experiencing very small probe field.}
\label{fig:Approaching}
\end{figure}

Fig.~\ref{fig:Approaching} shows the performance of the protocol with electron
spin resonance signals obtained in field gradients varying by more than three
orders of magnitude as probe-sample separation $d$ is decreased. Panel (a)
shows the signal at a probe-cantilever separation of 150 $\mu$m where $G_z \sim
30$ T/m. The MRFM spectral width $\Delta H$ is determined by both the field
inhomogeneity $G_z \Delta z$ across the sample width $\Delta z$ and the
inhomogeneity of the micromagnetic probe field. In this case, $G_z \Delta z
\sim 1.5 \, \rm G \ll \Delta H \sim 15$ G; rather the spectral width is mainly
due to the frequency modulation: \(2 \Omega_{\rm dev}/\gamma = 14\) G.

Fig.~\ref{fig:Approaching}(b) shows the evolution of the MRFM signal with
decreasing probe-sample separation. At smaller separations, signals occur at
lower applied field because the probe field experienced by the sample is
increasing. Plotted in the inset are the tip fields experienced by the
resonating spins, estimated by subtracting the external field values at which
the peak signals occur from the resonant field. They agree well with
calculations of the field from a 44 $\mu$m diameter uniformly magnetized NdFeB
sphere. As the separation decreases, the widths of the spectra increase. Whilst
the modulation deviation remains constant, $G_z$ increases a thousand-fold as
$d$ decreases from $150 \, \mu$m to $d = 2 \, \mu$m where \(G_z \sim 10^5\)
T/m. At small separations the signal widths are dominated by the field
inhomogeneity from the probe rather than the modulation deviation since $G_z
\Delta z \gg \Omega_{\rm{dev}} / \gamma$. As shown in figure
\ref{fig:Approaching}(c), the $\Delta H > 4$ kG in this gradient.

Despite the large increase in gradient, the maximum MRFM forces vary relatively
weakly:  from 0.07 to 0.3 pN throughout the range of separations. While the
force per spin is proportional to the gradient, the number of resonating spins
from an homogeneous sample is approximately inversely proportional to the
gradient. So, to first order, the total force is independent of the gradient.
Fig.~\ref{fig:Approaching}(c) also shows our single fitting parameter modeling
of the force spectrum at 2 $\mu$m. The two prominent features near the
resonance fields \( \omega_{\rm rf}/ \gamma = \pm 2.1\) kG are the
zero-probe-field resonances (ZPFR) arising from sample regions remote from the
probe magnet where its field and gradient are small \cite{h:jmr02}. The spectra
at $d=2$, 29 and 48 $\mu$m in Fig.~\ref{fig:Approaching}(b) and (c) are
one-dimensional spin-images \cite{zr:jap,sv,meier:mrfmlocalspc} of our sample
with varying spatial resolution ($\propto 1/G_z$), all obtained using the CASMO
protocol.

We have demonstrated a new protocol enabling spin-imaging with seamless
variable resolution---from high to low field gradient---that avoids
restrictions imposed by the cantilever response time. Frequency modulation
manipulates resonant spins to generate a periodic force on the cantilever
synchronized with its position.  This force is evident as a cantilever
frequency shift and hence can be detected with a bandwidth independent of
cantilever response time.  This phase control further allows phase-sensitive
lock-in detection and hence atto-Newton force detection \cite{r:singlespin,
r:90-nmresolutionimaging}. By expanding the range of measurement parameters
open to MRFM measurement and enabling variable resolution spin-imaging, this
moves MRFM closer to the goal of obtaining widely applicable imaging and
characterization tool.

This work was supported by the Army Research Office through MURI grant
W911NF-05-1-0414.

\end{document}